\documentclass[12pt,preprint]{aastex}
\begin{document}
\title{Information of Structures in Galaxy Distribution }
\author{Fan Fang}
\affil{Spitzer Science Center, California Institute of Technology, Pasadena, CA 91125}

\begin{abstract}
We introduce an information-theoretic measure, the R\'{e}nyi information,
to describe the galaxy distribution in space.  We discuss properties
of the information measure, and demonstrate its relationship with
the probability distribution function and multifractal descriptions.
Using the First Look Survey galaxy samples observed by the Infrared
Array Camera onboard Spitzer Space Telescope, we present measurements
of the R\'{e}nyi information, as well as the counts-in-cells distribution
and multifractal properties of galaxies in mid-infrared wavelengths.
Guided by multiplicative cascade simulation based on a binomial model,
we verify our measurements, and discuss the spatial selection effects
on measuring information of the spatial structures.  We derive structure
scan functions at scales where selection effects are small for the
Spitzer samples.  We discuss the results, and the potential of applying
the R\`{e}nyi information to measuring other spatial structures.
\end{abstract}


\section{Introduction}
\label{sec:intro}

The large-scale spatial distribution of galaxies is an important
topic for modern cosmology.  The cosmic structure as revealed by 
the observed galaxy spatial distribution is believed to originate
from primordial density fluctuations.  Gravitation amplifies
these fluctuations and is the main driver for the formation and
evolution of the cosmic structures.  In the current popular
scenario, galaxies form inside the previously collapsed ``dark''
gravitational wells in a process joined and modified by gas dynamics,
radiative cooling, and photonization.  The coalescence of these
dark halos brings galaxies together and to merge in a hierarchical
manner.

The large-scale distribution of the galaxies can be characterized
by various statistical and topological methods.  In particular,
the 2-point correlation function has been extensively used.  It
measures the second moment of the probability distribution, and
statistically completely describes a Gaussian density field, which is
believed to represent the primordial density fluctuations.  However, the
density field smoothed over the observed galaxy spatial distribution
is highly non-Gaussian.  The evolved cosmic structure as probed by
galaxy distribution contains highly dense regions crowded by galaxies
delineating spatial voids where few galaxies are located.  We generally
need the probability distribution function, or its moments, to
completely characterize such galaxy distribution in space.  A
counts-in-cells method has been used to establish the galaxy probability
distribution function.  Theory \citep{sh84, sf96} and models
\citep{cm83, fry84, fry85}
have been developed to interpret the probability distribution.

A multifractal description for galaxy spatial distribution has been
studied both theoretically \citep{piet87,jmse88,borg93} and
numerically \citep{vbp92}, and applied to several galaxy samples
\citep{mjdw90,bpv93}.  In particular, Borgani (1993) studied
the multifractal behavior of various hierarchical probability
distribution functions and derived the behavior of multifractal dimensions
for extreme underdense and overdense regions.  Indeed, the geometrical
concepts of fractal and multifractal are appealing given the ubiquitous
presence of such structures in various natural and social phenomena
\citep{mandelbrot83}.  Less-well-perceived has been the statistical
origin of multifractals as characterizing the moments of a probability
distribution.  For a review of multifractal applications in
large-scale structure, see Coleman \& Pietronero (1992) and Borgani (1995).

The purpose of this paper is to introduce R\'{e}nyi information
as a valid characterization of any spatial structure, including
galaxy distribution.  We show that R\'{e}nyi information, being closely
related to probability distribution and multifractal measures, probes
the statistical moments sensitive to any levels of under- and overdense
spatial structures.  At scales where the information contents are 
well-preserved and can be accurately quantified, statistical moments
are jointly described by R\'{e}nyi information and dimensions,
for which the underlying generator has a physical origin.  We also
illustrate the procedure by applying the R\'{e}nyi information, along
with the probability distribution and multifractal measures, to
observed galaxy samples in the infrared wavelengths as well as a
simulation.

In the next section, we introduce R\'{e}nyi information and the properties,
the relations to the moments of the probability distribution function
and to the multifractal measurements.  In Section \ref{sec:res}
we present the results of the probability distribution and
R\'{e}nyi information for the infrared samples observed by the Infrared Array
Camera (IRAC) onboard Spitzer Space Telescope.  We discuss a multiplicative
cascade simulation in Section \ref{sec:sim}, which provides means of
validating our methods of measurement, and showing the effects of spatial
selections in our galaxy samples.  We further derive the functions scanning
the structure of the moments for the samples
based on simulation results.  We discuss our results and potential
applications of the information measure in Section \ref{sec:dis}.

\section{R\'{e}nyi Information, R\'{e}nyi Dimensions, and Structure Scan Functions }
\label{sec:him}

Shannon \& Weaver (1948) derived an information measure to
describe the amount of information needed in order to know
the occurrence of an event with a given probability.  In an
important development, R\'{e}nyi (1970) expanded
Shannon's information measure to arbitrary orders.  Suppose
we have $N_{c}$ cells placed to cover a distribution of
$N_{g}$ galaxies.  This can either be a 2-dimensional angular
or 3-dimensional spatial distribution.  The probability
$p_{k}$ of finding a galaxy in a given cell $k$ containing
$N_{k}$ galaxies is $p_{k}=N_{k}/N_{g}$.  The R\'{e}nyi
information is defined as 

\begin{equation}
\label{eqn:ls}
I_{\beta} = \frac{1}{\beta-1}\log\sum_{k=1}^{N_{c}}p_{k}^{\beta},
\end{equation}
where $\beta$ is the information order which in principle can be
any real number (although in our application we consider integers only).
At positive orders the overdense structures dominate the information
estimate, whereas the underdense structures contribute the most to the
information measure at negative orders.  At $\beta=1$, the R\'{e}nyi
information reduces to the Shannon information.

The summation term for the probabilities $p_{k}$ to order $\beta$
can also be written as 

\begin{equation}
\label{eqn:re}
\sum_{k=1}^{N_{c}}p_{k}^{\beta} = \sum_{N_{i}=0}^{N_{g}}N_{c}f(N_{i})(\frac{N_{i}}{N_{g}})^{\beta},
\end{equation}
where $f(N_{i})$ is the galaxy probability distribution function.
Therefore the R\'{e}nyi information of order $\beta$ is related to
the $\beta$-moment of the probability distribution as

\begin{equation}
\label{eqn:infomo}
I_{\beta}=\frac{1}{\beta-1}[\log\frac{N_{c}}{N_{g}^{\beta}} + \log \sum_{N_{i}=1}^{N_{g}}N_{i}^{\beta}f(N_{i})], \hspace{5mm} \beta \neq 1,
\end{equation}
\begin{equation}
\label{eqn:infomo1}
I_{\beta}=\frac{N_{c}}{N_{g}}\sum_{N_{i}=1}^{N_{g}}N_{i}\log(\frac{N_{i}}{N_{g}})f(N_{i}), \hspace{5mm} \beta = 1,
\end{equation}
which is in turn related to the volume-averaged $\beta$-point correlation
function \citep{peebles80}.  The relation is intuitively easy to
understand as $\sum p_{k}^{\beta}$ is simply the total probability
of finding $\beta$ galaxies in a cell.  At positive orders of integral
$\beta$ the R\'{e}nyi information characterizes the amount of information
corresponding to the event of finding $\beta$ galaxies in the cells
covering the discrete galaxy distribution.

Some properties of the R\'{e}nyi information indicate the behavior
of the moments of the galaxy spatial distribution.  It can be
proved \citep{beck90} that

\begin{equation}
\label{eqn:cmp1}
(\sum p_{k}^{\beta1})^{1/(\beta1 - 1)} \leq (\sum p_{k}^{\beta2})^{1/(\beta2 - 1)}, \hspace{5mm} \beta1 < \beta2,
\end{equation}
\begin{equation}
\label{eqn:cmp2}
(\sum p_{k}^{\beta1})^{\frac{1}{\beta1}} \geq (\sum p_{k}^{\beta2})^{\frac{1}{\beta2}}, \hspace{5mm} \beta1 < \beta2 \hspace{5mm} \rm and \hspace{5mm} \beta1\beta2>0.
\end{equation}

Taking the logarithm we get

\begin{equation}
\label{eqn:cmp3}
I_{\beta1} \leq I_{\beta2}, \hspace{5mm} \beta1 < \beta2,
\end{equation}
\begin{equation}
\label{eqn:cmp4}
\frac{\beta1-1}{\beta1}I_{\beta1} \geq \frac{\beta2-1}{\beta2}I_{\beta2}, \hspace{5mm} \beta1 < \beta2 \hspace{5mm} \rm and \hspace{5mm} \beta1\beta2>0.
\end{equation}

Since $0 < p_{k} \leq 1$ we have
$\sum p_{k}^{\beta} \leq \sum p_{k}=1$ for $\beta>1$, and
$\sum p_{k}^{\beta} \geq \sum p_{k}=1$ for $\beta<1$.  Therefore
there is an upper limit $I_{\beta} \leq 0$ for all $\beta$.
We need zero information, or have perfect knowledge for an event when
$I_{\beta}=0$.  The bounds are also reflected by

\begin{equation}
\label{eqn:cmp5}
I_{\beta2} \leq \frac{\beta2}{\beta2-1}\frac{\beta1-1}{\beta1}I_{\beta1} \hspace{5mm}\rm for \hspace{5mm} 1<\beta1<\beta2 \hspace{5mm} \rm and \hspace{5mm} \beta1<\beta2<0,
\end{equation}
\begin{equation}
\label{eqn:cmp6}
I_{\beta2} \geq \frac{\beta2}{\beta2-1}\frac{\beta1-1}{\beta1}I_{\beta1} \hspace{5mm} \rm for \hspace{5mm} 0<\beta1<\beta2<1.
\end{equation}

The R\'{e}nyi information depend on the cell size $l$, and diverge as
$l \rightarrow 0$.  One property that remains finite at this limit is
the so-called R\'{e}nyi dimensions:

\begin{equation}
\label{eqn:rd}
D(\beta)=\lim_{l \longrightarrow 0}\frac{I_{\beta}}{\log l}.
\end{equation}

Any galaxy distribution becomes discontinuous
at the scale of the typical galaxy separation.  The above limit is
not achieved in a discrete distribution or in practical measurements.
A more practical definition for galaxy distribution is the ``effective''
R\'{e}nyi dimensions, for which we calculate the slope of $I_{\beta}$ versus
$\log l$.  There is no reason \it a priori \rm to expect the slope for
a given order to be a constant over all scales for a given structure.
In fact, this is not implied in equation \ref{eqn:rd} for a continuous
multifractal distribution.  We call it a simple multifractal if the
effective R\'{e}nyi dimension for any given order has a single slope
across all scales.

Examining R\'{e}nyi information and dimensions over information orders
is identical to inspecting the structure of statistical moments of
a distribution.  Studying such scan functions has the advantage
of summarizing infinite amount of parameters (moments) in just a
few relations for a statistical distribution.  Here we relate R\'{e}nyi
dimensions to a scan function defined in a continuous multifractal
field.  Suppose $\langle\epsilon_{l}\rangle$ is the field density measured and
ensemble-averaged at scale $l$.  Function $K(\beta)$ is the scaling
exponent for moments of the field (also called the structure function)
$\langle\epsilon_{l}^{\beta}\rangle \propto l^{-K(\beta)}$.
Now that the R\'{e}nyi dimension is practically $D(\beta)=dI_{\beta}/d\log l$,
since $\langle\epsilon_{l}^{\beta}\rangle \propto \langle N^{\beta}\rangle l^{-D\beta} \propto \sum p_{k}^{\beta}l^{-D\beta+D}$,
where $N$ is the counts in the cells of size $l$, $D$ is the dimension of
the space in which the distribution is embedded (e.g. $D=2$ in our
applications below), we obtain

\begin{equation}
\label{eqn:cod}
K(\beta) = (\beta-1)(D-D(\beta)).
\end{equation}

Function $K(\beta)$ is therefore also called the codimension \citep{sl87}.
Here we use a general name, the structure scan function, for the R\'{e}nyi
information and dimensions as functions of $\beta$, as well as for functions
like $K(\beta)$.

Th multifractal dimensions are usually defined by using the generalized
correlation integral \citep{hp83, gp83}.  Many measurements of the multifractal
properties of galaxy distribution have been based on measuring the generalized
correlation integral, which uses cells of varying sizes centered at selected
galaxies.  Such a procedure is not valid for estimating
R\'{e}nyi information since neighboring cells bound to cross each other
above a certain scale.  Below this scale there is a non-zero probability that
some of the galaxies are not covered by the ensemble of cells.  Either case
changes the normalization for the probabilities, and the R\'{e}nyi information
is not accurately quantified for the original structure.  This is further
explained in Section \ref{sec:sim}.  We want to
emphasize here that \it not only the slope of the R\'{e}nyi information versus
scale (R\'{e}nyi dimensions), but also the R\'{e}nyi information
itself is a physical measurement, both being open to physical interpretations.
\rm We will further discuss this point in Section \ref{sec:dis}.

We note that the differential form of the second-order
correlation integral is called the conditional density, which had been used to
characterize galaxy distribution in early surveys \citep{cps88, ls91}.

The R\'{e}nyi dimensions basically show the scaling properties
of R\'{e}nyi information.  Loosely speaking, a multifractal galaxy
distribution has a position-dependent scaling exponent $\alpha(x)$ in
$p_{k}\sim l^{\alpha}$.  It can be shown strictly \citep{schuster95}
that the spectra of these scaling exponents $f(\alpha)$ and the R\'{e}nyi
dimensions (multiplied by a factor of $(\beta-1)$) are related by a
Legendre transformation

\begin{equation}
\label{eq:spect}
f(\alpha) = -\tau(\beta) + \beta\alpha.
\end{equation}
where $\tau(\beta)\equiv(\beta-1)D(\beta)$, $d\tau/d\beta=\alpha$, and
$df/d\alpha=\beta$.  A number of interesting properties of $D(\beta)$
and $f(\alpha)$ are discussed in Beck (1990).  A few interesting
ones include $D(\beta)$ being a decreasing function of
$\beta$ and bounded as $\beta \rightarrow \pm\infty$,
$D(+\infty) = \alpha_{min}$, and $D(-\infty) = \alpha_{max}$.
These limits and the ways $D(\beta)$ and $f(\alpha)$ approach the
limits show the properties of the moments of the spatial distribution
from a scan function perspective.

\section{Measurements and Results }
\label{sec:res}

The IRAC instrument onboard the Spitzer Space Telescope
provides fresh view into the cosmos in the mid-infrared
wavelengths of 3.6 $\mu$m, 4.5 $\mu$m, 5.8 $\mu$m, and 8.0 $\mu$m.
The Spitzer First Look Survey (FLS) using IRAC provides
a uniform coverage of a 4 square-degree field centered at
RA $=17^{h}18^{m}$, Dec $=59\arcdeg30\arcmin$ with
a total 60-second exposure time for each pixel in the $256 \times 256$
arrays \citep{lacy06}.  For our present purpose, we use the full galaxy samples
established for an earlier 2-point correlation analysis \citep{fang04}
across the IRAC wavelengths.

We divide the two-dimensional area covered by an IRAC sample into square
cells of varying sizes.  The cells are non-overlapping and contiguous
for the purpose of accurately estimating the R\'{e}nyi information.
The boundary of the sample area and the usage of a cell are determined
by the mask files used to establish the galaxy sample \citep{fang04}.
We always have $>500$ ``good'' cells at the largest scales of measurement
to ensure good statistics.  At smaller scales the cell numbers are
much greater.

For each sample and each cell size we count the number of galaxies in the
cells and establish the histograms in Figure \ref{fig:cic}.  The histograms
represent the estimates of the probability distribution function, from
which the moments of the distribution can also be measured.  For each histogram,
we plot the fit of the theoretical Gravitational Quasi-equilibrium Distribution
Function (GQED) \citep{sh84, sf96}.  The single fitting parameter $b$,
the average ratio of the gravitational correlation potential energy to
twice the kinetic energy, is also shown in the plots.  For comparison,
we also draw the Poisson distributions with the same mean galaxy counts
in cells of given sizes.  Apparently the galaxy distribution
deviates more from the Poisson distribution at larger cell sizes,
indicating the effect of galaxy clustering in IRAC wavelengths.
The GQED, on the other hand, describes the distributions of IRAC
galaxies remarkably well over all scales.

We follow the same procedures as in the counts-in-cells experiment
to divide the IRAC sample areas into square cells to calculate
the R\'{e}nyi information.  Figure \ref{fig:infos} shows the
relation between the cell sizes and the measured R\'{e}nyi
information at orders from $1$ to $20$.  The R\'{e}nyi information
scales with cell sizes, but the relation is not linear for our
galaxy samples.  We will discuss below the effects that can potentially
change the scaling relation.  The apparent crowding of the curves
at high information orders implies an upper limit, which is $0$ based
on the above discussion, for the information measures.  The limit
constrains the behavior of the moments of galaxy distribution
in the information space.

Intuitively, exclusion of structures, such as galaxies not covered
by the cells, or regions that cells avoid due to masking, changes the
information content of the structure.  Although we intend to cover the
sample using contiguous cells, changing cell-size causes some galaxies
in the sample not being covered by cells of a new size due to the sample
boundary and masked areas.  Until these effects can be fully
accounted for, we are in fact measuring the information of slightly
different structures at each scale, even though the probability is
normalized by the total number of galaxies covered by cells.  This
introduces noises in the information measurements.  In the next section,
we study these effects using simulation of a known structure.

\section{Simulation and Further Results }
\label{sec:sim}

To verify our results, we generate a multiplicative cascade simulation
based on a binomical model.  The binomial model was found to describe
well the multifractal scaling in the dissipation field of fully developed
turbulence \citep{ms87}.  We use the binomial model for its simplicity
and analytically derivable relations for R\'{e}nyi information and
multifractal properties.  The multiplicative cascade method was
formulated to study energy transfers at different scales in turbulence
\citep{mandelbrot74,frisch95}.  It is by far the most effective method
to simulate a multifractal field.

We use a discrete multiplicative cascade simulation, consistent with our
purpose to study counts at multiple length scales.  The simulation aims
to create distributions of counts at ten different scales within a given
area with a conserved overall number density $D_{0}$.  At the first level,
the area is divided into four quadrants of the same size, two of which
hold a fraction $p/2$ of $D_{0}$, and the other two have the fraction
$(1-p)/2$ (so the overall number density is conserved as $D_{0}$; also
called the canonical process).  There are many ways to distribute the
two number densities equally among four cells.  We choose to have a fixed
pattern of distribution here.  We tested different patterns and the results
are the same.

At the second level, each quadrant area is further divided into four
identical (smaller) cells, with the distribution of the same probabilities
of the same pattern.  The number counts in a cell is the product
of the probability assigned at this level multiplied by the probability
(of the quadrant covering the cell) at the previous level (and by an
arbitrary total source number, which we choose to use $1$).

We continue the process to generate smaller and smaller cells and their
number counts.  We stop at the tenth level where we have data over 10
scales (of ratio 2) for statistics.  At level $n$, a cell has the number
count proportional to $(\frac{p}{2})^{k}(\frac{1-p}{2})^{n-k}$, where $k$ is
an integer between $0$ and $n$.  Therefore it is called the binomial model.
The resulting structure, although modulated sharply by cell edges, is a
simple multifractal.  Based on Halsey et al. (1986) and Meneveau \& Sreenivasan (1987), we derive
the R\'{e}nyi information, the R\'{e}nyi dimensions, and the spectra of the
multifractal scaling exponent for the 2-dimensional binomial field as

\begin{equation}
\label{eqn:binfo}
I_{\beta}=\frac{n}{\beta-1}\log[2(\frac{p}{2})^{\beta}+2(\frac{1-p}{2})^{\beta}],
\end{equation}

\begin{equation}
\label{eqn:mdbi}
D_{\beta}=1+\frac{1}{1-\beta}\log(p^{\beta}+(1-p)^{\beta}),
\end{equation}
 
\begin{equation}
\label{eqn:idbi}
f(\alpha) = 1+\log(\frac{\log(1-p)-\log(p)}{\log(1-p)+\alpha})+\frac{\log(p)+\alpha}{\log(1-p)-\log(p)}\log(\frac{\log(p)+\alpha}{-\alpha-\log(1-p)}),
\end{equation}
where $n$ is the level number in the cascade (e.g. smallest scale being
$n=10$), and $\beta$ is the information order.

Figure \ref{fig:infosim} shows the scaling of R\'{e}nyi information 
in the binomial field.  The measurements, using the same methods and
algorithm used for the infrared samples, are indicated by points in
the Figure.  The lines are calculated based on equation \ref{eqn:binfo}.
The agreement is nearly perfect for all orders.

Next we compare measurements of the R\'{e}nyi dimensions and multifractal
spectra with those predicted by equations \ref{eqn:mdbi} and \ref{eqn:idbi}.
Since the scaling of the measured R\'{e}nyi information in Figure
\ref{fig:infosim} is well-represented by lines, we use linear least-square
fit for each information order in that figure to obtain the R\'{e}nyi
dimensions.  We confirm from the fit that the information values
at the smallest scale (where $\log r=0$) and the slope are both
within $10^{-3}\%$ of those predicted by equations \ref{eqn:binfo}
and \ref{eqn:mdbi}.  The slope values from
the fit for each order are plotted as dots in the left-panel of
Figure \ref{fig:spectsim}.  The line is the scan function for R\'{e}nyi
dimensions based on equation \ref{eqn:mdbi}.  To obtain the multifractal
spectra, we use a cubic spline fit to model the measured R\'{e}nyi
dimensions and derive the $\alpha$
and $f(\alpha)$ values within the range of information orders.
These values are again represented by the dots in the right-panel
of Figure \ref{fig:spectsim}.  The line in the Figure is based on
equation \ref{eqn:idbi}.  In both panels the values based on the measured
R'{e}nyi information agree well with the predicted ones.
This shows the reliability of the methods and algorithms we use.

In Figure \ref{fig:corsim} we show measurements of generalized correlation
integral superimposed on the plane of R\'{e}nyi information versus scale
for our simulation.  We generate $1000$ cell positions in the
$1024 \times 1024$ field, and vary the size of the cells centered around
these positions.  Cells may overlap, but are ignored if they cross the
field boundary.  The generalized correlation integral is calculated using
the remaining cells, following the standard algorithm \citep{mjdw90}.
In the Figure the dash-lines are generalized correlation integral
measurements, and solid-lines are the predicted R\'{e}nyi information.
It is clear that although the R\'{e}nyi dimensions may be approximately
maintained, the generalized correlation integral does not measure R\'{e}nyi
information.  We also find that the values of the generalized correlation
integral depends on the number of cells in the experiment, further 
strengthening this point.

To investigate the effects of spatial selection of structures, we include
the mask files on which the IRAC samples are based.  Each of the mask files
is a FITS representation of the FLS field with a dimension of
$6200 \times 6600$ pixels.  Since our simulation has a different dimension,
we first project these masks onto a $1024 \times 1024$-pixel field.
This procedure maintains the scale ratio of the masked areas and the
field size.  We then follow the same criteria to exclude cells in the
simulation field overlapping with projected masked areas, and repeat
the procedures of measuring the R\'{e}nyi information in the simulated field.

The results, using the mask files for the four IRAC samples, are shown
in Figure \ref{fig:masksim}, with solid-line predictions
superimposed on measured points connected by dotted lines.  There is an
obvious effect of spatial selection on measuring both R\'{e}nyi information
and dimensions.  Notably involving the IRAC masks introduces an apparent
scale-dependency of R\'{e}nyi dimensions, particularly at greater scales, where
both the R\'{e}nyi information and dimensions are higher than predicted.
At smaller scales, there is a systematic offset to higher (negative)
information values, although the slopes for R\'{e}nyi dimensions are
approximately maintained.  Masking reduces the amount of structures
in the original binomial field, and smaller amount of information
(identical to the absolute value of the measured R\'{e}nyi information)
is needed to know an event occuring (such as $\beta$ sources in a cell)
with a given probability.  Cells of increasing scales may cover
a fluctuating, but generally increasingly smaller samples from the original
structure accompanied by a mask.  This confirms our intuition that
R\'{e}nyi information is an intrinsic property of a spatial structure.
Any modifications of the structure modify its information content.
Other derived properties such as the R\'{e}nyi dimensions can also be
affected if not measured properly.  While the geometry and pattern of
the four mask files vary, the effects on the R\'{e}nyi information are
remarkably similar.

Although it changes the information contents of the original structure,
it appears that the IRAC masks preserve the scaling of the information
at smaller scales.  The graininess of galaxy distribution, however,
introduces a Poisson limit for these smaller scales (remember our
simulation is not grainy), below which a cell contains either one
or no galaxy in most of the regions.  Any multifractal behavior breaks
down at this limit.  At scales smaller, the number of cells contributing
to the R\'{e}nyi information is roughly identical to the total number of
galaxies, and the R\'{e}nyi information reaches a (lower) limit also and
flattens out (also see Figure \ref{fig:structinfo} below).  Both these
effects at large and small scales can make the R\'{e}nyi information
curves of a multifractal become concave.  This curvature is observed in
Figure \ref{fig:infos}.

Based on the IRAC sample sizes and the un-masked areas for the samples,
we estimate the mean separations of any two galaxies in the samples,
which are roughly $20"$ for IRAC-1 and 2, $55"$ for IRAC-3 and 4 samples, 
assuming uniform distributions.  We use these as the lower scale limit
for reliable multifractal estimate.  For upper limit,
Figure \ref{fig:masksim} implies a linear scale of $\sim 1\%$ of the
field size, assuming that the scale ratio applys to the FLS field.
This is only slightly higher than the lower limit of IRAC channel-3
and 4 samples.  Basically, the smaller number of galaxies in these samples
combined with the amount of masking prevented us from reliably estimating
the multifractal behavior for these two samples.

For illustration purpose, we perform a cubic spline fit to each of the
R\'{e}nyi information relations in Figure \ref{fig:infos}, and derive
the R\'{e}nyi dimensions and the scan function
for IRAC-1 and 2 samples at a scale of $45"$.  We perform another cubic
spline fit to the scan functions (like we did for the binomial field)
and obtain $\alpha$ and $f(\alpha)$
throughout the range of R\'{e}nyi dimensions.  In Figure \ref{fig:spect}
we show these relations for the two FLS samples.  The figure illustrates
how R\'{e}nyi dimensions decrease with increasing information order, and
converge to a limit.  Also $f(\alpha)$ appears to be a convex function
of $\alpha$.  Where $f(\alpha)=0$, the $\alpha$ values represent the
R\'{e}nyi dimension limit when $\beta\rightarrow\infty$.  All are
typical behaviors of multifractals.

In Figure \ref{fig:structinfo}, we plot the R\'{e}nyi information as
a function of order, a different type of scan function, measured for
orders $-20$ to $20$ at scales of $20"$, $32"$, $44"$, $55"$,
and $68"$ for all IRAC samples.  At most of these scales the masking
effect is small, where the information can be measured accurately for
galaxy distribution.  For samples at IRAC channels 3 and 4, however,
Poisson effects dominate the three smaller scales.  This is told by
the scan curves converging to the Poisson limit below or near information
order zero.  For all IRAC samples, the limit is shown by the scan curve
behavior at negative information orders, where the information measure
is sensitive to and dominated by underdense regions in the samples.
At positive orders and at scales where information can be measured accurately,
the scan curves tell the structures of high moments of the galaxy distribution.

\section{Discussion }
\label{sec:dis}
 
We have shown that the R\'{e}nyi information, the effective R\'{e}nyi
dimensions, their structure scan functions, and the multifractal spectra
contain the properties of the high moments of a spatial distribution. 
These measurements can be used to scan properties of these high moments.
These properties detect the amount of deviation from Gaussian densities,
and are highly constrained in the parameter space in these measurements.

Our experiments also show that spatial selection effects are important
and can bias these measurements.  Any selection modifies the original
structure and the amount of R\'{e}nyi information the structure contains.
Depending on the amount of selection, the R\'{e}nyi dimensions may be
maintained over a limited range of scales above the Poisson limit for discrete
distributions.  One needs to conduct controlled experiments such as simulations
to verify at these scales.  For IRAC-1 and 2 samples, there is indication
in Figure \ref{fig:infos} that the information-scale relation is still
not linear within the range.  It is yet uncertain how much of this is
caused by masking as well as by approaching the Poisson limit, both effects
leading systematically to a concave curve, or if there is scale-dependency
for the R\'{e}nyi dimensions in our IRAC samples, which would imply
a more complex structure than a simple multifractal distribution
at these scales.

Whether galaxy spatial distribution is a multifractal, or whether
homogeneity can be reached at large scales, as cosmological principle
states, has been observationally a controversial issue
\citep{peebles93, cp92, avnir98, martin99}.  Our analyses show
that caution needs to be exercised extrapolating a multifractal
structure to small and large scales, particularly if spatial selection exists
for a galaxy sample, even if multifractality is observed at
scales more reliable for multifractal measurements.

It may be possible to recover the lost information in a galaxy sample
by ``filling-in'' the masks based on known properties of galaxy distribution.
Such known properties may come from minimally-masked samples of
galaxies of the same type, or from $N$-body simulations, for examples.
Just as the $\delta$-function for generating the probability in our
multiplicative cascade simulation, there is a variety of statistical
functions that can serve as the generating functions for simulating
full-scale multifractal fields \citep{gw93}.  Among these generating
functions the log-L\'{e}vy distribution is of particular interest due to
the unique position of the L\'{e}vy distribution in replacing a Gaussian
in the generalized central-limit theorem where variances of the component
distributions can be infinite, and also due to its applications to
a ``universal class'' of geophysical structures \citep{sl87}.
The structure scan functions are uniquely determined by probability generating
functions which are of physical origin.  The generating function
would be a significant property to know if galaxy distribution is a
multifractal to large scales.

Another way to seek physically interpreting the R\'{e}nyi information is
to use the moments of the probability distribution via equations
\ref{eqn:infomo} and \ref{eqn:infomo1}, which are not restricted to a
multifractal structure.  Since $N_{c}\propto l^{-D}$, where $D$ is the
dimension of the space the distribution is embedded, we can also derive
for equation \ref{eqn:infomo}

\begin{equation}
\label{eqn:dm}
D_{\beta} = -\frac{D}{\beta-1} + \frac{1}{\beta-1}\frac{d\log m_{\beta}}{d\log l},
\end{equation}
where $m_{\beta}$ is the $\beta$-moment of the probability distribution
function.  It is clear from the relation that we have a simple multifractal
distribution across all scales only if $d\log m_{\beta}/d\log l$ is not
a function of scale.  This is not the case for GQED, for example.
On the other hand, any physically-derived probability distribution can
interpret the R\'{e}nyi information and dimensions via these relations.

Independent of the multifractality of a structure, the R\'{e}nyi
information and dimensions are general characterizations of statistical
properties of the structure.  A simple multifractal is a special and very
restrictive type of structure in its practical definition.  The R\'{e}nyi
information and dimensions and their corresponding scan functions
can describe any types of structures, whether or not multifractals.
The R\'{e}nyi information is extensive, whereas its scaling, or
``information rate'' with changing scales is an intensive parameter.
Both are important for a given structure.  As we collect galaxy samples
from surveys with greater area coverage and increasing depth as well as
in more wavelength channels, we are collecting increasingly more information
about the large-scale structure, and the absolute values of the measured
R\'{e}nyi information increase at a given scale.  Any variations
of the R\'{e}nyi dimensions, on the other hand, are of different origin.

For spatially confined structures, such as a giant molecular cloud,
the extensivity of R\'{e}nyi information also depends on resolution.  A more
resolved observation reveals more detailed structure, and therefore
more information contents.  While the R\'{e}nyi information
and dimensions can be identically applied to continuous and discrete
spatial fields, it is important to recognize what properties
are used for measurement.  It is clear that we want to characterize the
moments of a spatial structure, and that we can use spatial densities
for measuring.  An astronomical observation is usually a radiation
measurement, however, and the proportionality between the two is
only an assumption.  For non-astronomical structures, the meaning
of the measurements can be more clear-cut.

The R\'{e}nyi information and dimensions can also be applied to
one-dimensional time-series.  In the temporal domain, the amount of
time-delay serves as scaling, and the information contents and rate
describe the temporal structure built by distributions of the change
of the observed properties over certain and different time-spans.
An information measure is a measure about the knowledge of a structure
or system, and therefore its predictability.  It would be desirable to
quantify the predictability of a statistical distribution or a time-series
using R\'{e}nyi information and dimensions.  So far research on this
topic remains limited.

The relation between the R\'{e}nyi information and dimensions measured in
2-dimension and those in 3 dimensional space for the same structure
can be straightforward.  The 2-dimensional cells used to cover a
structure can also be 3-dimensional cells with the third dimension extended
to cover the same structure.  When properties such as spatial density
can be accounted for by measurements when projecting the structure onto
a 2-dimensional area, the information is not lost.  The only uncertainty
is the correspondence between the 2-dimensional and 3-dimensional scales.
It is, however, a generally interesting question what scales are measured
by cells of non-identical dimensions.  For galaxy spatial distribution,
the evolutionary effects of galaxies in the third dimension need to be
disentangled from projection before the structure can be analyzed
in three dimensions.

\acknowledgments 

I thank the anonymous referee for providing constructive comments.
I thank the FLS team at the Spitzer Science Center for assembly
of the FLS data products.  The Spitzer Space Telescope is
operated by the Jet Propulsion Laboratory (JPL), California Institute of
Technology under NASA contract 1407.  Support for this work was provided
by NASA through JPL.



\clearpage

\figcaption[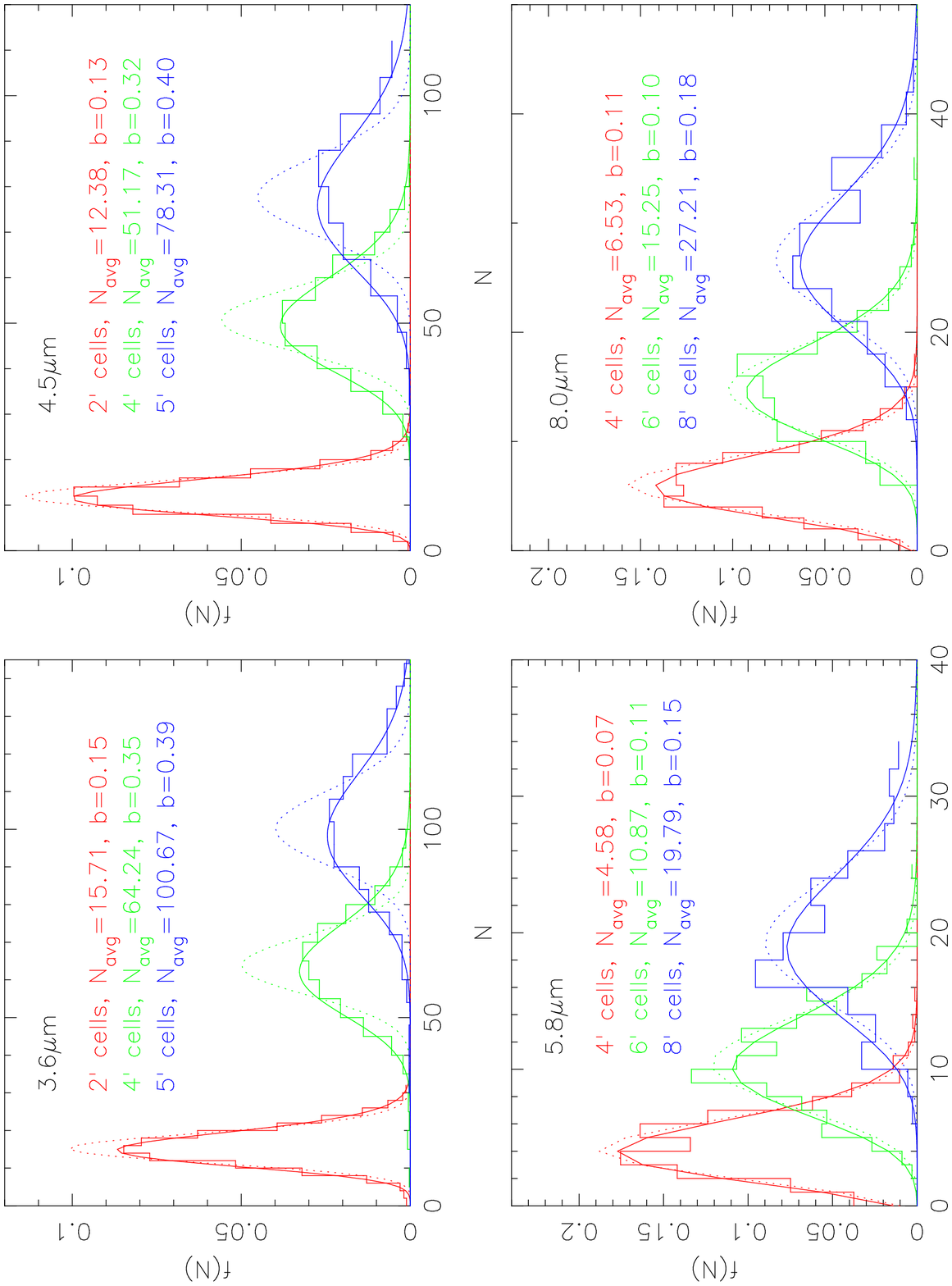]{
The 2-dimensional counts-in-cells distribution of the IRAC galaxies.
Square Cells of three different sizes on the side, indicated in the
Figure, are used for each IRAC channel data.  Histograms show the counts
of the IRAC sources.  The solid lines are the fit using the Gravitational
Quasi-equilibrium Distribution Function; fitting parameter $b$ is shown
in the Figure for each case.  The dotted lines are the Poisson
distributions with the same average number of galaxies $N_{avg}$
in given size cells as in the corresponding histograms.
\label{fig:cic}}

\figcaption[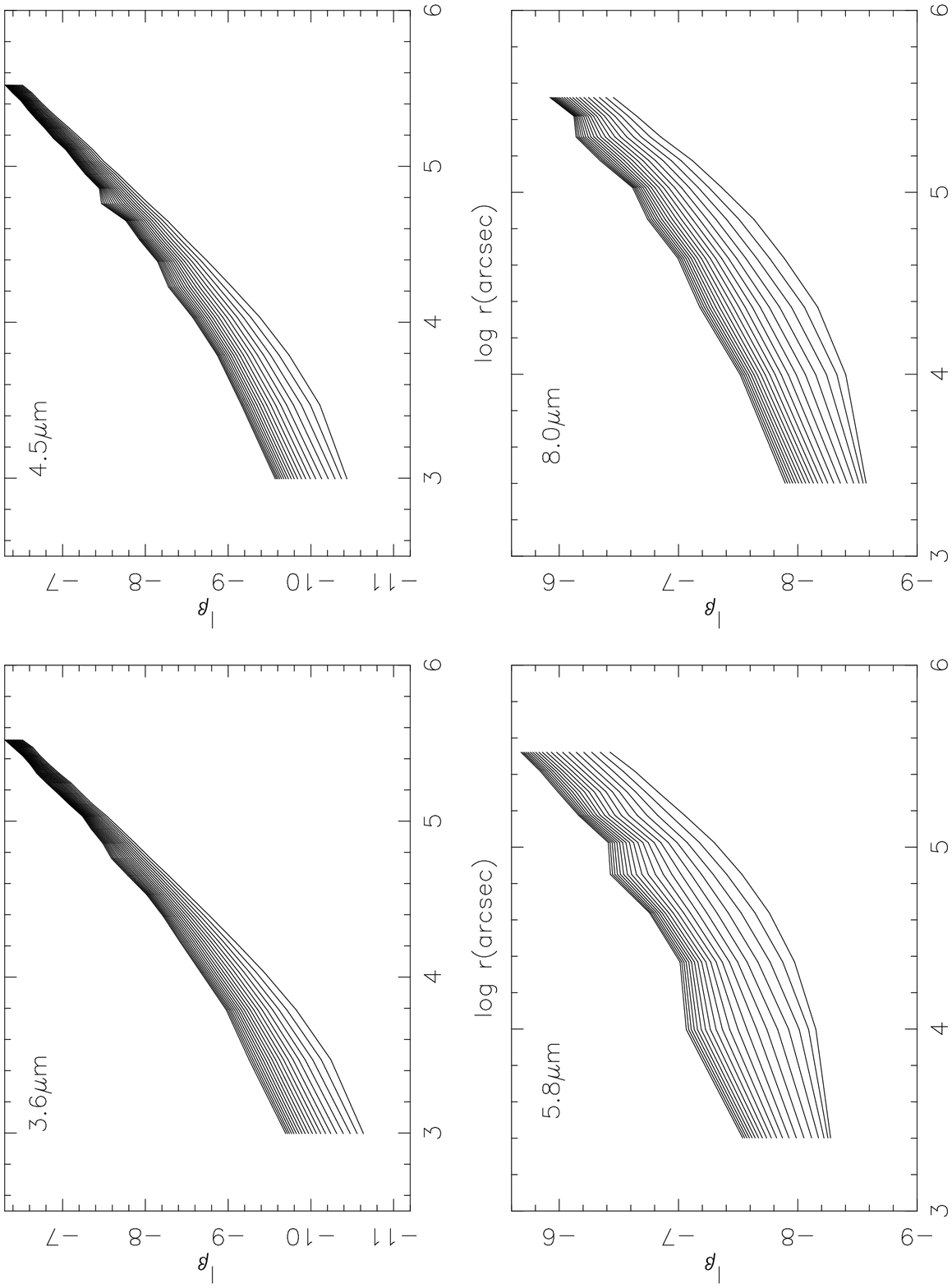]{
Relation between R\'{e}nyi information and the cell sizes used to measure
the information.  Each curve represents an information order.  For each order
we measure the information at different cell sizes ranging from 20 to
250 arcsec for IRAC-1 and 2 samples, and from 30 to 250 arcsec for IRAC-3
and 4 samples.  The measurements are connected by lines.  The orders are
from 1 to 20 from bottom to top.  The R\'{e}nyi information are in ``bits''
with $\log 2$ as the unit.
\label{fig:infos}}

\figcaption[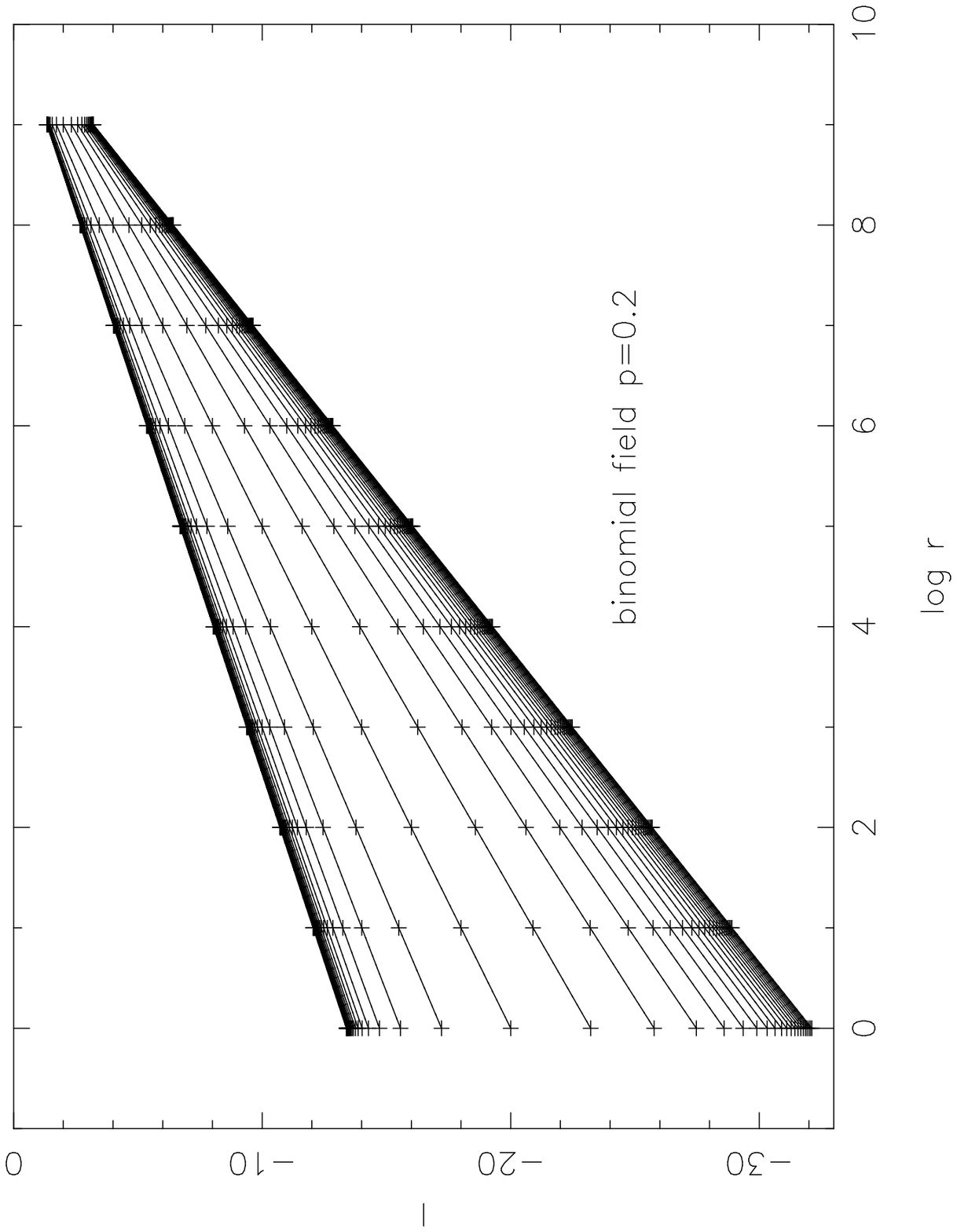]{
R\'{e}nyi information for orders $-20$ to $20$ (bottom to top) as a function
of scale in the simulated binomial field with $p=0.2$.  The crosses are
measurements using the same algorithm as for Figure \ref{fig:infos}.  The lines
are predictions of the binomial model with $p=0.2$.  The logarithmic base is $2$.
\label{fig:infosim}}

\figcaption[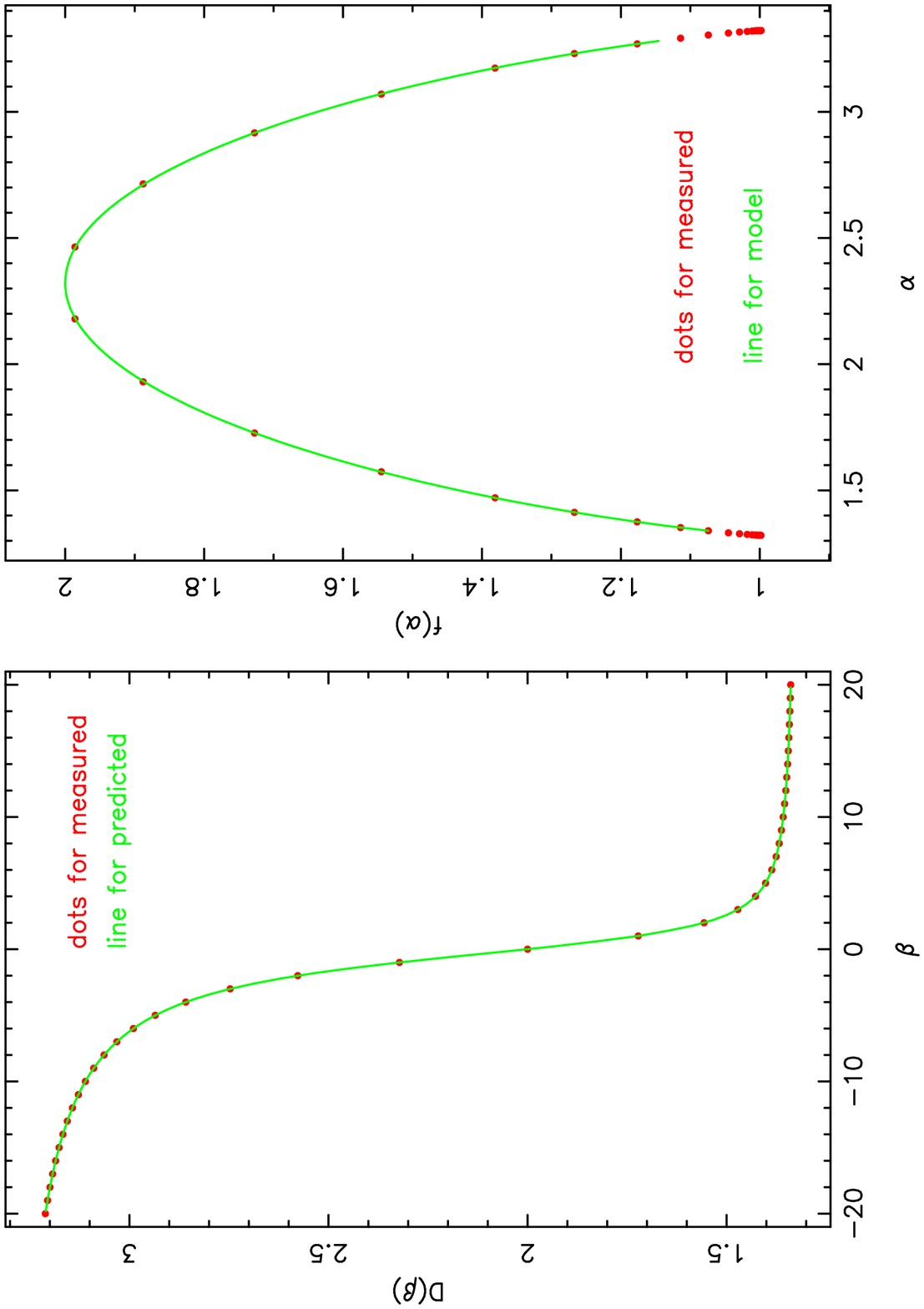]{
R\'{e}nyi dimensions as a function of information order and the spectra
of scaling exponents in the simulated binomial field.  The dots are
measurements, and the lines are predictions based on the binomial model.
\label{fig:spectsim}}

\figcaption[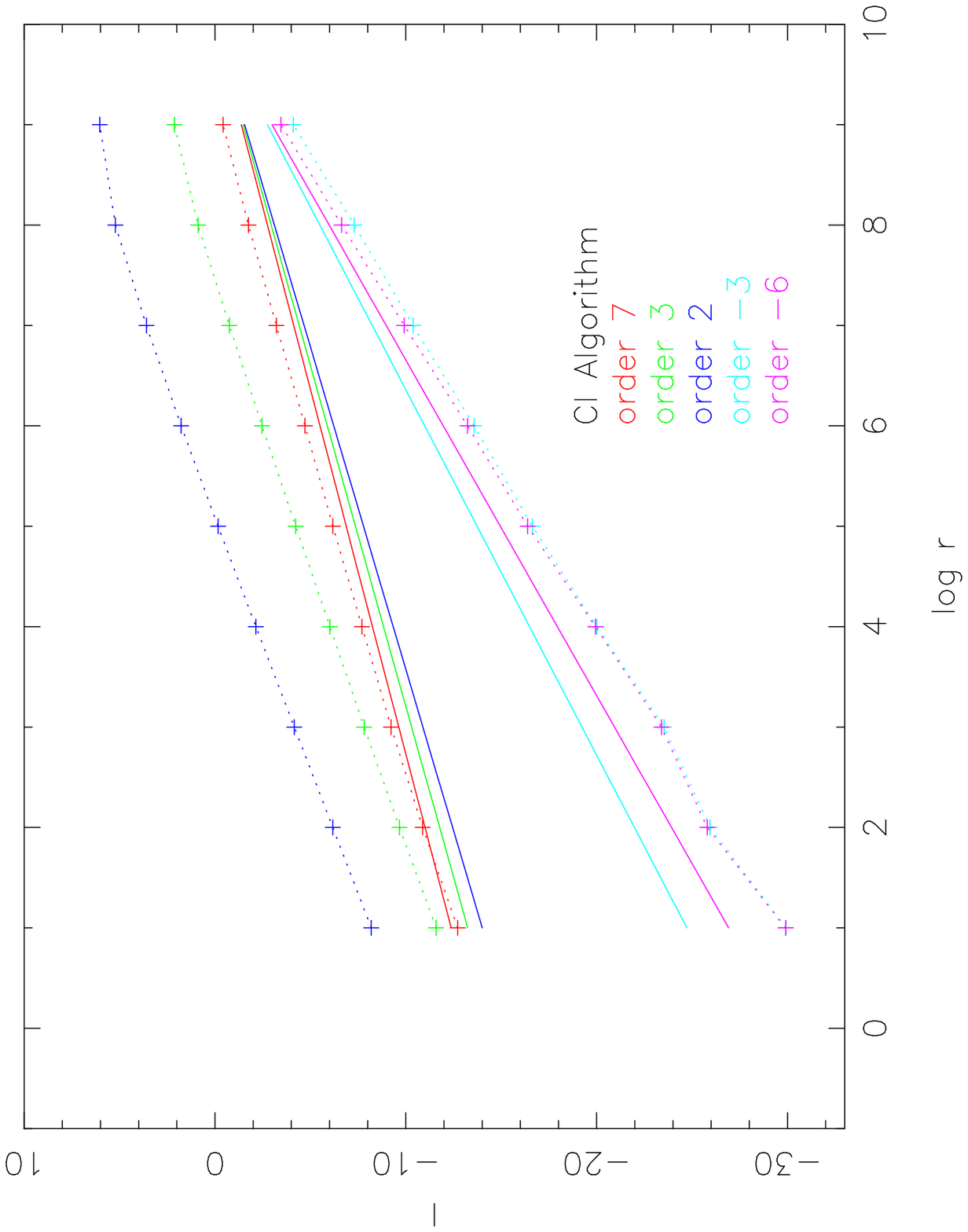]{
Generalized correlation integral measurements in the same binomial-field
simulation.  A total of $1000$ cell positions are distributed in the
$1024\times 1024$ grid area.  The cells are centered at these positions with
their sizes vary for each measurement.  The values of the generalized
correlation integral are measured at nine scales for information orders
-6, -3, 2, 3, and 7, shown by the dashed lines.  Solid lines are the
predicted R\'{e}nyi information at these scales.  Here the logarithmic
base is $2$.
\label{fig:corsim}}

\figcaption[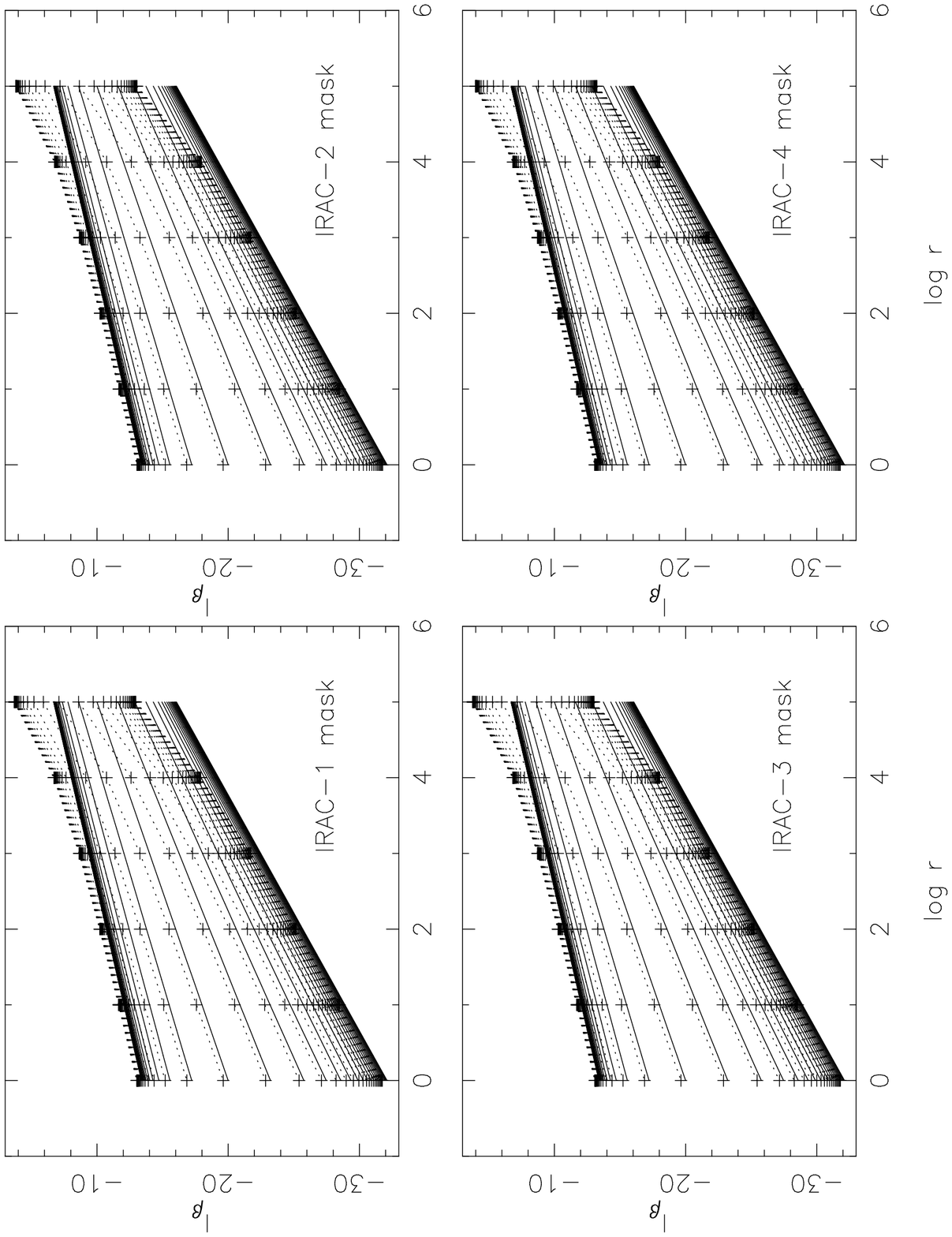]{
R\'{e}nyi information for orders $-20$ to $20$ (bottom to top) as a
function of scale in the simulated binomial field modified by FLS
IRAC masks scaled to the size of the simulation field.  The crosses
are measurements, connected by dotted lines.  The solid lines are
predictions as in Figure \ref{fig:infosim}.  The logarithmic base
is $2$ for the figures.
\label{fig:masksim}}

\figcaption[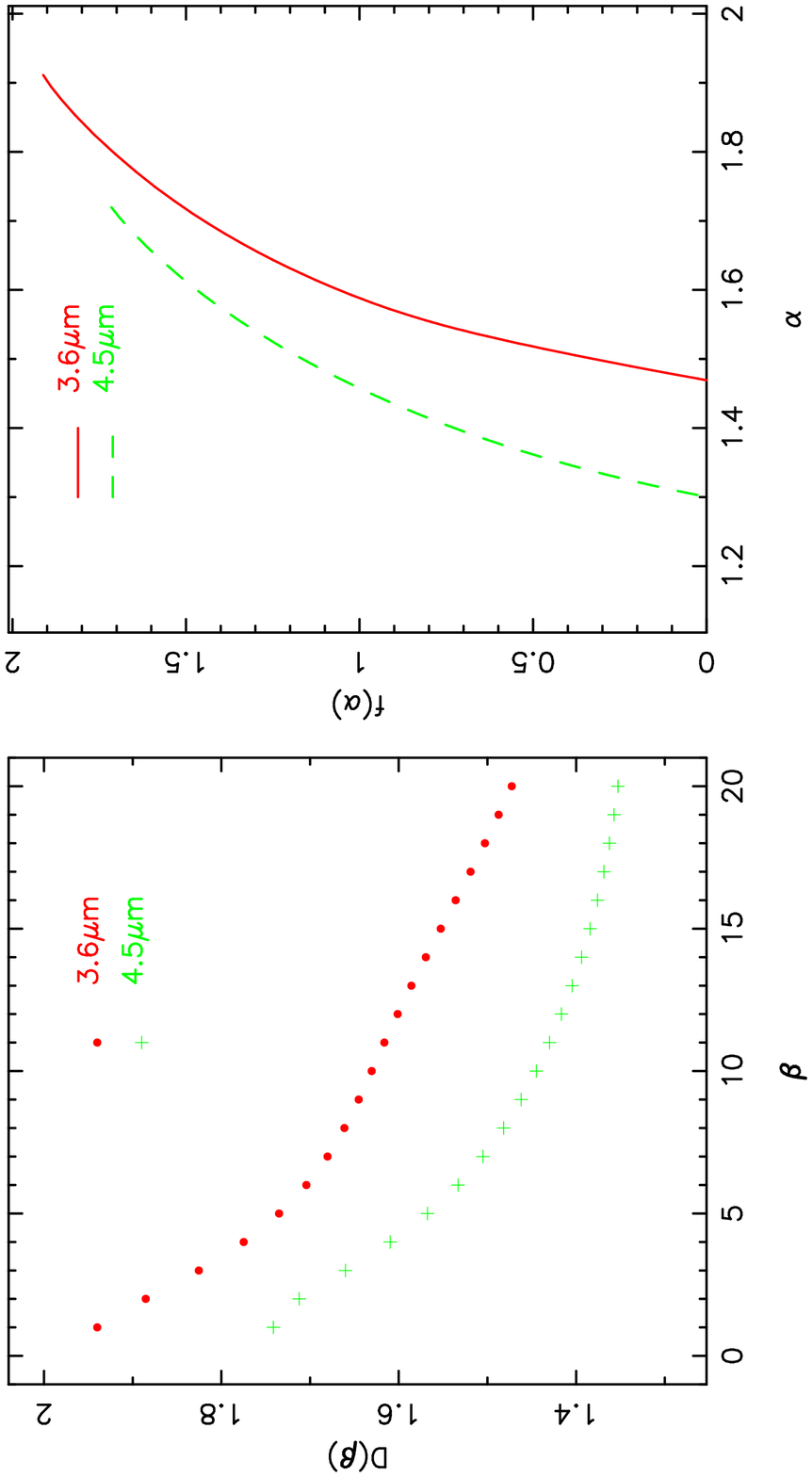]{
R\'{e}nyi dimensions as a function of information order and the spectra of
scaling exponents, estimated at a scale of $45"$ for IRAC channel-1 and 2
galaxies.  A cubic spline fit is performed for each curve in Figure
\ref{fig:infos} for estimating the R\'{e}nyi dimensions at different
information orders.  This generates the left-panel of the figure for the
scan function.  A second cubic spline fit is used to estimate the scaling
exponents and their spectra from the scan function, which produces the
right-panel figure.
\label{fig:spect}}

\figcaption[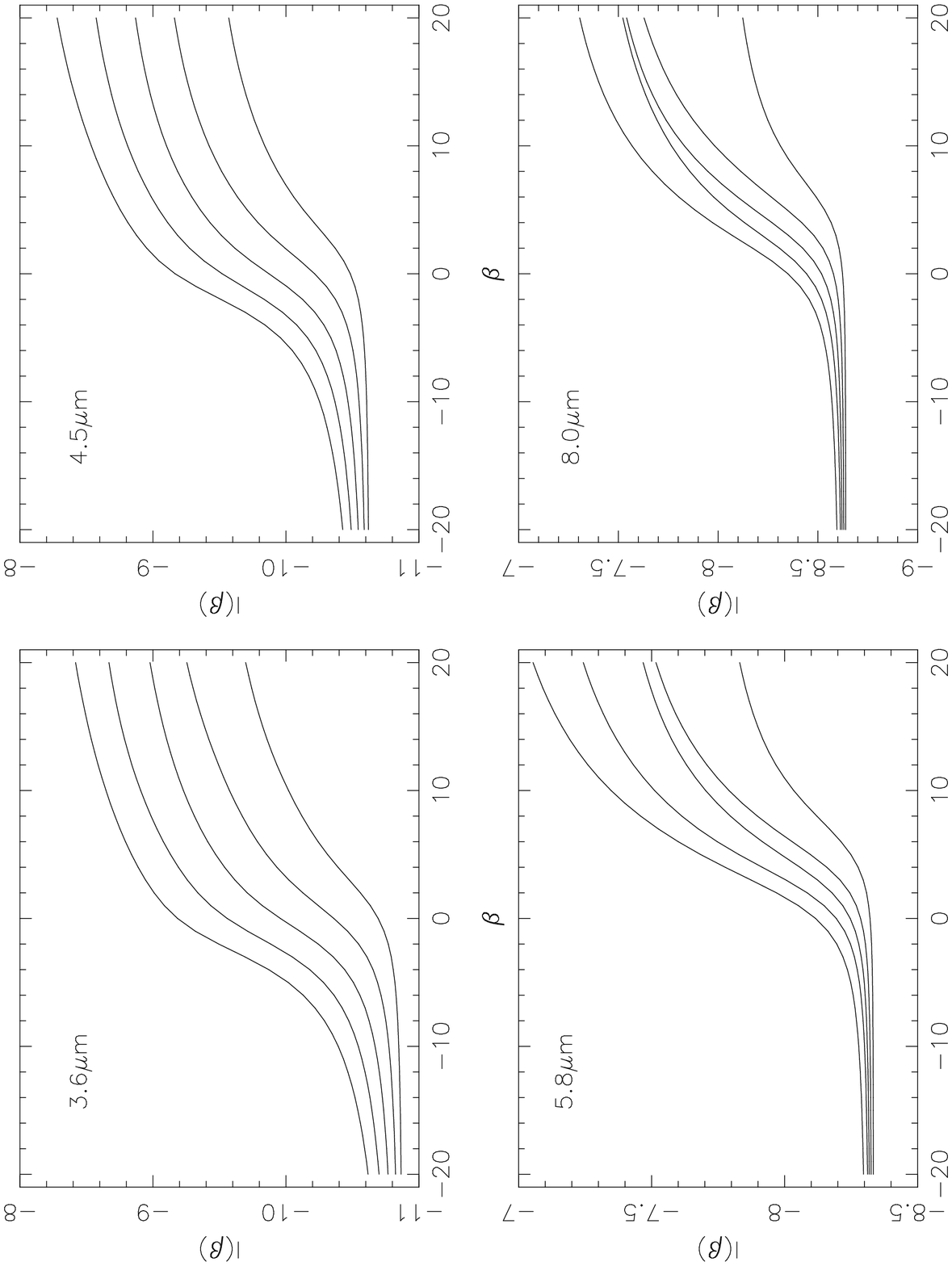]{
Structure scan functions of the R\'{e}nyi information.  From top to bottom
the lines represent the R\'{e}nyi information measured at scales of
$68"$, $55"$, $44"$, $32"$, and $20"$ for orders from $-20$ to $20$.
At nagative information orders where underdense structures dominate,
the graininess of galaxy distribution leads the R\'{e}nyi information
to the Poisson limit.
\label{fig:structinfo}}

\newpage
\plotone{fig1.eps}
\newpage
\plotone{fig2.eps}
\newpage
\plotone{fig3.eps}
\newpage
\plotone{fig4.eps}
\newpage
\plotone{fig5.eps}
\newpage
\plotone{fig6.eps}
\newpage
\plotone{fig7.eps}
\newpage
\plotone{fig8.eps}

\end{document}